\documentclass[epjCONF]{svjour}
\usepackage{graphics}
\usepackage[varg]{txfonts} 
\usepackage[latin1]{inputenc}
\usepackage{lscape}
\usepackage{graphicx}
\usepackage{rotating}
\usepackage[square]{natbib}
\bibpunct{[}{]}{,}{n}{}{,} 
\newcommand{\aap}{A\&A}
\newcommand{\aapr}{A\&A Rev.}

\newcommand{\aj}{AJ}
\newcommand{\apj}{ApJ}
\newcommand{\apjl}{ApJ}
\newcommand{\apjs}{ApJS}
\newcommand{\apss}{Ap\&SS}
\newcommand{\araa}{ARA\&A}
\newcommand{\cosbold}{{\tt CO$^5$BOLD}}
\newcommand{\gca}{Geoch. Cosmoch. Acta}
\newcommand{\msait}{MSAIt}
\newcommand{\mnras}{MNRAS}
\newcommand{\nat}{Nature}
\newcommand{\npa}{Nucl. Phys. A}

\newcommand{\prl}{Phys. Rev. Lett.}
\newcommand{\rppp}{Rep. Progr. Phys.}
\newcommand{\solphys}{Sol. Phys.}
\newcommand{\ssr}{Space Sc. Rev.}

\session-title{Ageing low mass stars: from red giants to white dwarfs}
\begin{document}
\title{Selected topics in the evolution of low-mass stars}
\author{M. Catelan\inst{1,2}\fnmsep\thanks{\email{mcatelan@astro.puc.cl}} }
\institute{Departamento de Astronom{\'\i}a y Astrof{\'\i}sica, 
Pontificia Universidad Cat{\'o}lica de Chile, Av.\ Vicu{\~n}a Mackenna 4860, 782-0436 Macul, Santiago, Chile. 
\and The Milky Way Millennium Nucleus, Av.\ Vicu{\~n}a Mackenna 4860, 782-0436 Macul, Santiago, Chile.} 
\abstract{
Low-mass stars play a key role in many different areas of astrophysics. In this 
article, I provide a brief overview of the evolution of low-mass stars, and discuss
some of the uncertainties and problems currently affecting low-mass stellar models. 
Emphasis is placed on the following topics: the solar abundance problem, 
mass loss on the red giant branch, and the level of helium enrichment associated 
to the multiple populations that are present in globular clusters. 
} 
\maketitle
\section{Introduction}\label{sec:intro}
Low-mass stars are the most common type of stars in the Universe, and as such 
a proper understanding of their properties is of great astrophysical importance. 
The structure and evolution of low-mass stars has been the subject of several recent 
reviews \citep[e.g.,][and also the papers by M. Salaris and A. Weiss in these 
proceedings]{aw06,mc07,sc11,sc12,gvea12}~-- and therefore, in the present paper, 
I will only briefly discuss a few selected 
topics of current interest. I start by providing an overview of low-mass stellar 
evolution in \S\ref{sec:over}. A personal view of the current status of the so-called 
``solar abundance problem'' is given in \S\ref{sec:sun}, and some recent developments in 
the computation of the mass-loss rates of red giants 
are described in \S\ref{sec:m-dot}. Some remarks 
on the helium content of the multiple populations that are present in globular clusters 
(GCs) are provided in \S\ref{sec:he}. Finally, conclusions are given in \S\ref{sec:conc}.

\section{Overview}\label{sec:over}
\begin{figure}[!htbp]
  \centerline{
  \includegraphics[angle=0,scale=0.532]{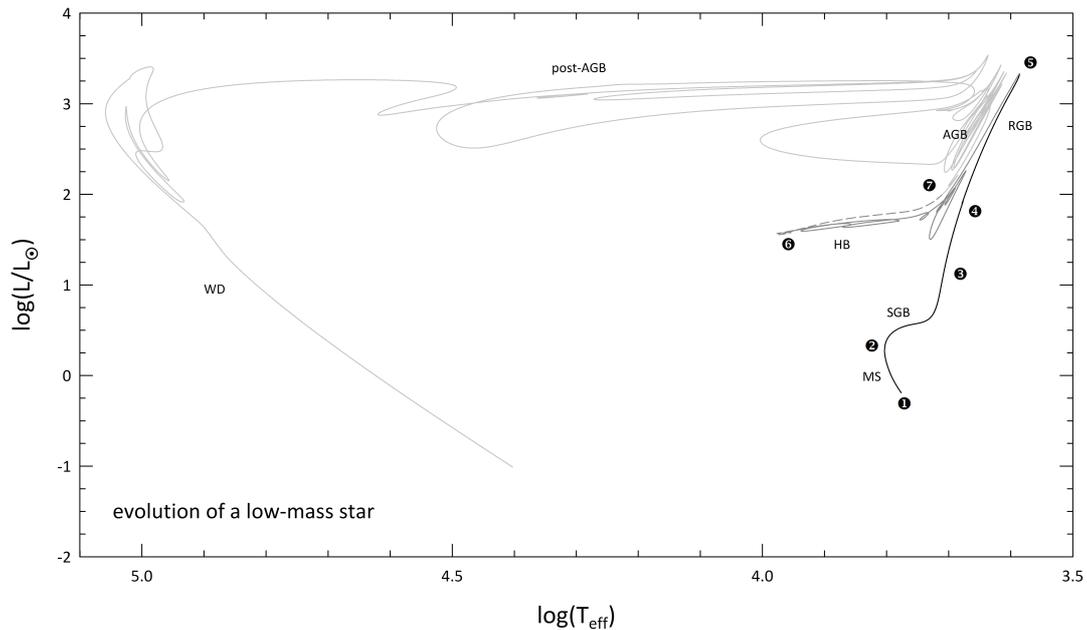}}
  \caption{H-R diagram showing the evolution of  a low-mass star with the following 
  parameters: $Y = 0.23$, $Z = 0.0015$, $M = 0.862 \, M_{\odot}$. The star loses a total
  of $\approx 0.26 \, M_{\odot}$ due to stellar winds during the RGB phase.   
  The main evolutionary stages are labeled, and 
  key stages in the star's life are numbered as follows: (1) ZAMS; 
  (2) central H exhaustion and turn-off point; (3) first dredge-up episode; (4) RGB
  ``bump''; (5) He ignition (RGB tip); (6) ZAHB; (7) central He 
  exhaustion. Adapted from \cite{mc07}, based on computations by \citet{tbea01}.     
  }  
  \label{fig:track}
\end{figure}

\begin{figure}[t]
  \centerline{
  \includegraphics*[width=3.25in]{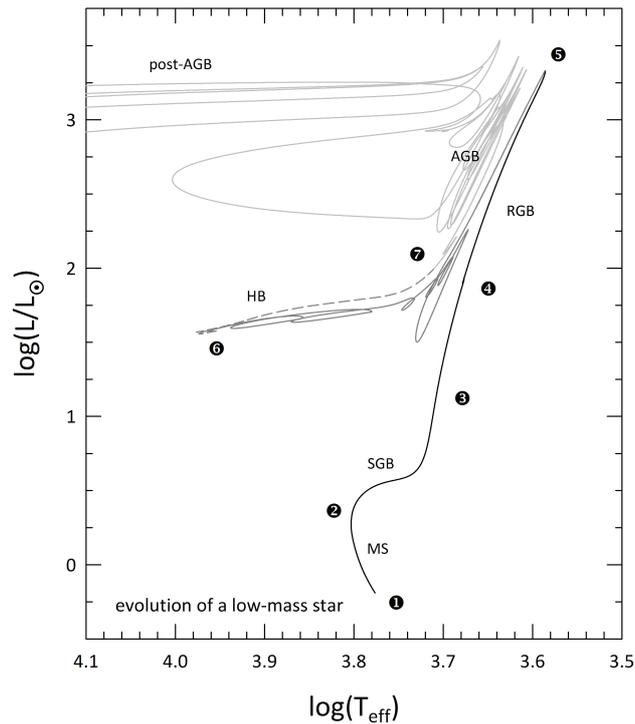}}
  \caption{As in Figure~\ref{fig:track}, but zooming in on the main nuclear burning stages. 
   }
      \label{fig:track-zoom}
\end{figure}

The evolution of a ``typical'' (single) low-mass star accross the Hertzsprung-Russell
diagram, from the zero-age main sequence (ZAMS) to the white dwarf (WD) cooling curve, 
is shown in Figure~\ref{fig:track}. All the main evolutionary phases are highlighted, 
including the MS, subgiant branch (SGB), red giant branch (RGB), horizontal branch (HB), 
asymptotic giant branch (AGB), post-AGB, and WD phases. In addition, several key points 
in the evolution of the star are also labeled, namely: 1) ZAMS; 2) core H exhaustion/turn-off
point; 3) first dredge-up episode; 4) RGB ``bump''; 5) He core ignition under degenerate
conditions (He ``flash''); 6) zero-age HB (ZAHB); 7) core He exhaustion/commencement of 
the AGB phase. A zoomed-in version of this plot is provided in Figure~\ref{fig:track-zoom}; 
in this figure, the many gyrations or loops associated with thermonuclear flashes between 
the RGB tip and ZAHB are clearly seen, as are the loops associated with thermal pulses 
in AGB stars. Note that the star may cross the ``classical'' instability strip at several
points in its life, including not only the HB (RR Lyrae stars) and AGB (type II Cepheids) 
phases, but also the pre-ZAHB phase (``RR Lyrae stars in disguise'' \cite{vsaea08}). 
All these episodes in the life of a low-mass star (including, in addition, the pre-MS 
evolution) are extensively reviewed in 
\cite[][and references therein]{mc07}, and will thus not be repeated here.

\section{The Solar Abundance Problem}\label{sec:sun}

\subsection{The Problem }
\begin{figure}[t]
  \centerline{
  \includegraphics*[width=4in]{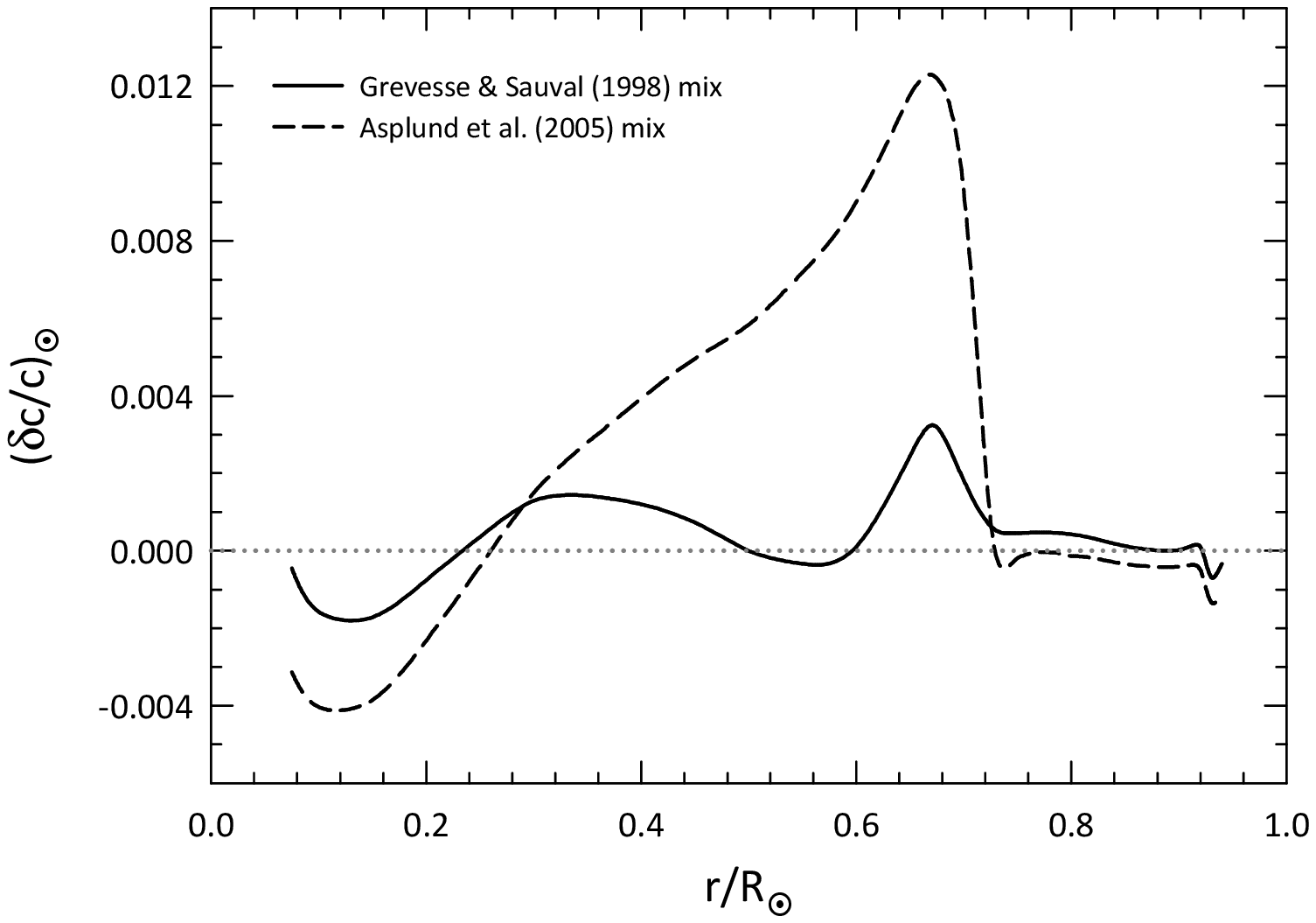}}
  \caption{Illustration of the ``solar abundance problem'' circa 2005. The relative 
    difference $(\delta c)/c$ between the sound speed as inferred from helioseismology and that 
    predicted by the ``standard solar model'' is shown for two different choices 
    of the solar heavy element mix: \citet[][{\em solid line}{\rm}]{gs98} and
	\citet[][{\em dashed line}{\rm}]{maea05}. 
   }
      \label{fig:ssm}
\end{figure}

\begin{figure}[!htbp]
  \centerline{
  \includegraphics*[width=4in]{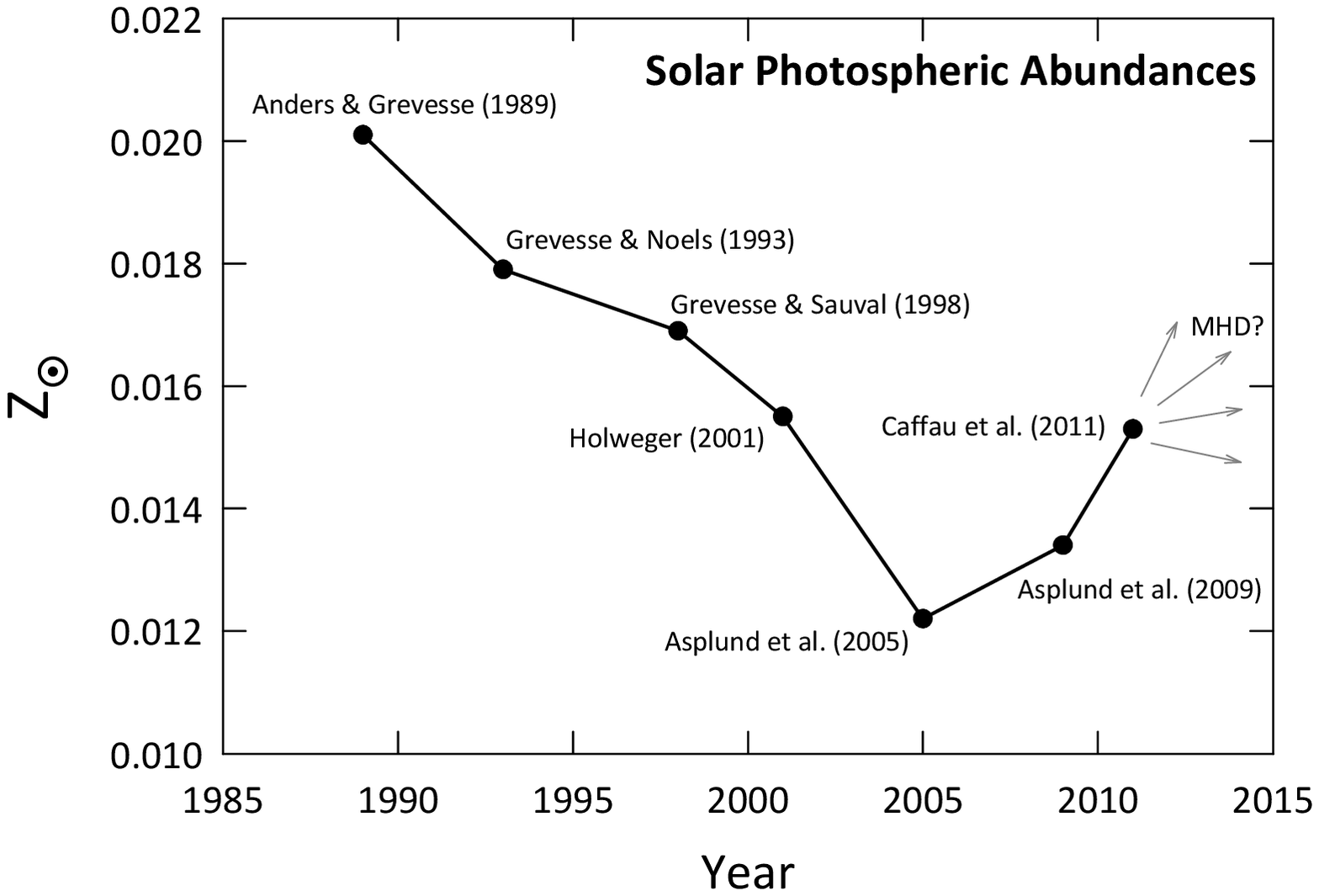}}
  \caption{Evolution of the recommended solar metal abundance $Z_{\odot}$. 
    The gray arrows point to the possible increase in $Z_{\odot}$ due to the inclusion 
	of MHD effects, as recently suggested by \citet{dfea10,dfea12}. 
   }
      \label{fig:solar-z}
\end{figure}

Until about 2005, it was generally considered that evolutionary computations based 
on the solar metallicity $Z_{\odot}$
favored by N. Grevesse and co-authors \citep{ag89,gn93,gs98}
``yield solar models in good agreement with the data'' \cite[][and references therein]{jbea05}. 
The perceived good agreement is illustrated by the solid line in Figure~\ref{fig:ssm}, which 
compares the predictions of the standard solar models computed by \citet{jbea06} for the 
sound speed $c(r)$ and the actual sound speeds that are inferred from helioseismology. The 
maximum deviation is of the order of 0.3\%. However, in 2005 new spectroscopic 
solar abundances were 
provided by \citet{maea05}, based on impressive, highly sophisticated 
3D hydrodynamical models of the solar 
atmosphere, which led to an important decrease in $Z_{\odot}$ (see 
Fig.~\ref{fig:solar-z}).\footnote{This plot includes the $Z_{\odot}$ value estimated 
by \citep{stcea04} on the basis of the \citet{hh01} abundance determinations for a few key 
elements, namely C, N, O, Ne, Mg, Si, Fe, as reported on Table~1 of \citet{gm10}.} 
Again as shown in Figure~\ref{fig:ssm}, this led to a significantly worse agreement between 
the solar model computations and the helioseismological observations, with the maximum 
deviation now reaching a level of about 1.2\%. 

This relatively recent development is commonly referred to as the ``solar abundance 
problem.'' To be sure, and as can 
be seen in Figure~\ref{fig:solar-z}, the $Z_{\odot}$ value went back up slightly with 
the extensive follow-up study by \citet{maea09}, but this increase proved insufficient 
to restore the level of agreement that existed previously \citep[e.g.,][]{asea09}.

The solution to the problem is currently unknown, but several suggestions have been raised
in the literature, none of which has proven entirely satisfactory. In what follows, a brief, 
and admittedly rather personal, 
overview is presented of some of the recent work that has been carried out in this area.

\subsection{Astrophysical Perspective}

\subsubsection{Solar Abundances: The Present and the Future} 
Recently, and as also shown in Figure~\ref{fig:solar-z}, 
the independent 3D hydrodynamic study by \citet{ecea11}, carried out 
using the \cosbold\ code, has brought a further increase in the solar metallicity, 
beyond that attained in \citet{maea09}. However, the \citeauthor{ecea11} results 
have recently been subject to some criticism, particularly in regard to the selection 
of atomic line lists \citep[][Grevesse 2012, these proceedings]{ngea11}. Accordingly, 
a change in the \cosbold\ $Z_{\odot}$ value cannot be excluded, at this point in time. 
However, it is unclear whether the change would bring values back down to the 
\citeauthor{maea09} levels. In particular, the controversy regarding the choice of 
line lists applies primarily to the carbon lines, and to a lesser extent to lines
of other elements (Ludwig \& Caffau 2012, priv. comm.). 

On the other hand, the possibility of yet another {\em increase} in $Z_{\odot}$ has 
been raised very recently by 
\citet{dfea10,dfea12}. Indeed, according to their recent 3D radiation-magnetohydrodynamic 
(MHD) computations, magnetic fields may play a very important role in deriving reliable 
abundances for the chemical elements in the Sun. Indeed, it has been known for several 
years now that the Sun has a substantial amount of ``hidden'' magnetic energy, in the 
form of tangled subsurface magnetic field lines, concentrated in intergranular lanes 
\citep{ikea12}, with an average intensity of the order of 100-200~G \citep{jtbea04}, 
but including contributions also at the kG level \citep{js12}. This line of research 
should certainly be pursued further, but at present it remains unclear by how much 
(and even whether) such 3D MHD models will increase $Z_{\odot}$ beyond current levels. 
MHD effects are accordingly illustrated in Figure~\ref{fig:solar-z} as gray arrows 
with uncertain slope. 

Recently, an increase in the Ne opacity has been raised as a
possible solution to the solar abundance problem 
\citep[see][for a recent review]{sb09}. 
However, the possibility of an increased Ne abundance
has been disputed by \citet{maea09}, and \citet{chlea07} show  
that increased Ne abundances actually {\em worsen} the agreement with the Sun, in 
the deeper part of the convection zone.

\subsection{Physical Perspective}
From a physical perspective, the situation immediately reminds one of that which was 
encountered a few decades ago 
in the field of stellar pulsation, with a persistent disagreement between 
computed and observed period ratios of double-mode (or ``bump'') Cepheids. \citet{ns82}
realized that the problem could be solved by increasing radiative opacities by a factor
of 2-3, at temperatures $\sim 10^5$~K. At the time, some authors considered that such a large
increase in the opacity would be unrealistic \citep[e.g.,][]{nmea84}, but 
\citeauthor{ns82}'s suggestions were eventually vindicated by the new opacity
calculations by both the OPAL and OP teams \citep[e.g.,][]{ciea87,ri92,msea94}. 
One may thus conjecture that further changes in the radiative opacities will occur
in the future that will bring helioseismology back into agreement with a low $Z_{\odot}$. 

Indeed, several authors have shown that an increase in the opacity by perhaps 10-30\%,  
in the temperature range $(3.5 \pm 1.5) \times 10^6$~K, and particularly close to the 
base of the convective zone, plus a more modest increase close to the center, 
would serve to at least restore the previous level of 
agreement between solar models and helioseismology 
\citep[e.g.,][]{ba04,jmea04,jbea05,sb09,jcdea09,as10,fv10,dg12}. 
However, it is widely thought \citep[e.g.,][and references therein]{jgea06} that 
the level of uncertainty in current opacity calculations is at the level of 5\% 
only \citep[but see][for a different point of view]{tcea11}. 
It thus remains to be seen whether 
such suggested opacity increases will materialize in the future, or whether alternative
explanations for the solar abundance problem will have to be found. In this sense, 
\citet{ya05} call attention to  
an alternative route towards increasing the ``effective'' opacity by means of internal 
gravity waves \citep{pr81}, which could provide another means of bringing solar models
back in agreement with helioseismology.  

While radiative opacities may be the most natural culprits, they are by no means the only 
possible ones; for instance, \citet{sb09,sb10} points out that changes in opacity cannot be 
fully satisfactory without accompanying changes in the equation of state.

\subsubsection{Non-Standard Solar Models} 
Many authors have explored the possibility that standard solar models may be  
missing some fundamental physical processes, whose inclusion might  
help bring them into agreement with helioseismology. Among the possibilities
that have been discussed in the literature, one may find the following: 
increased early mass loss \citep{gm10}; 
accretion of low-$Z$ material from the protosolar nebula \citep{gm10,asea11}; 
convective overshooting \citep{gm10}; 
realistic treatment of convection \citep{daea10}; 
enhanced diffusion \citep[e.g.,][and references therein]{jgea05,sb09}; 
rotation-induced mixing \citep{abea99,stcea04,mcea07}; 
and combinations thereof \citep[e.g.,][]{stcea11b}. 
Among the more exotic solutions, one may find 
the conversion of photons to new light bosons in the solar 
photosphere \citep{avea12}, as well as alternative theories 
of gravity \citep{jcea12}.

\subsection{An ``Effective Solar Metallicity'' for Evolutionary Databases}
At this point, it is worth reminding the reader that every set of evolutionary models 
currently in use by the astrophysical community at large 
is {\em calibrated on the Sun}. The procedure 
was pioneered by \citet{dl64} and \citet{dp64}, and requires that the present-day 
luminosity (or effective temperature) and radius of the Sun are matched with the 
present-day (i.e., at an age of 4.57~Gyr), precisely measured solar values. In this 
way, one obtains the solar helium abundance $Y_{\odot}$ and the mixing length parameter 
$\alpha_{\rm MLT}$ of mixing length theory, {\em for an assumed solar metallicity}: 
for instance, in the original work by \citeauthor{dl64}, a solar metallicity 
$Z_{\odot} = 0.03$ was adopted. This procedure has remained essentially unchanged 
over the past (almost) 50 years, and so present-day stellar evolution calculations, 
and the extensive sets of evolutionary tracks, isochrones, and luminosity functions
derived therefrom, and applied to studies of resolved and unresolved, Galactic and 
extragalactic, stellar populations, are still 
calibrated in much the same way~-- the main difference being that the adopted value
of the solar metallicity has changed with time (see Fig.~\ref{fig:solar-z}). Note 
that the very same procedure is always adopted, irrespective of the degree of physical 
refinements that are included or lacking in any set of models~-- for instance, 
gravitational settling, radiative acceleration, turbulent mixing, rotation, magnetic 
fields, treatment of convection beyond the extremely simplified MLT, etc. 

It is perhaps advisable, at this point in time, to recognize the limitations both 
of our input physics and models, and to calibrate these models using an additional  
quantity that was not available to the pioneers in the field, but which 
is now very precisely measured~-- namely, the {\em internal sound speed
profile} $c(r)$, as obtained from helioseismology \citep[e.g.,][]{sbea00,jbea06}. 
Solar neutrino rates can also be included in such a ``beefed up'' 
calibration process, thus providing 
yet another route to help further constrain the innermost layers of low-mass 
stellar models \citep[e.g.,][]{fv10,asea11,tcc11,vaea12,whea12}. 
Indeed, one is now in a priviledged position, from the standpoint of the available  
empirical data, to require of one's solar models a closer match not 
only of the Sun's luminosity and 
radius at an age of 4.57~Gyr, but also of $c(r)$~-- and especially the depth of the 
convection zone~-- in addition to the observed neutrino rates. 
To achieve this, an ``effective solar metallicity'' $Z_{\odot,{\rm eff}}$ may 
accordingly be defined in such a way that an optimum match 
of the present-day, precisely measured properties of the Sun~-- namely, its 
luminosity, radius, $c(r)$, and neutrino rates~-- yields $Y_{\odot}$, $\alpha_{\rm MLT}$, 
and $Z_{\odot,{\rm eff}}$. 


Ideally, in the presence of perfect physical input, perfect spectroscopic abundances, 
and perfect numerical models, $Z_{\odot,{\rm eff}}$ will be identical to the 
spectroscopically measured photospheric solar abundance $Z_{\odot,{\rm sp}}$. In 
practice, however, due to (not unexpected!) imperfections in any of these ingredients, 
calibration of the solar model will frequently give 
$Z_{\odot,{\rm eff}} \neq Z_{\odot,{\rm sp}}$. 

Whatever the explanation of the solar abundance problem, it is clear that  
solar models computed using the \citet{maea05,maea09} or \citet{ecea11} solar 
abundances fail to correctly predict the sound speed profile that is inferred from 
helioseismology. The fact that models computed using higher solar metallicities,  
as had previously been favored in the literature \citep[e.g.,][]{ag89,gn93,gs98}, 
provide a better match to the helioseismological observations 
strongly suggests that either the new solar metallicities are too low, or/and the 
physical ingredients that are used in the computation of stellar models, {\em and 
especially those that depend strongly on the metallicity} (such as radiative 
opacities), are somehow in error. Be as it may, {present-day standard solar models 
computed with low metallicities cannot be seen as properly describing the inner 
structures of low-mass stars, even if the low solar metallicity is correct. 
Evolutionary models (and the isochrones derived therefrom) calibrated on the Sun 
in the more robust way just described, 
on the other hand, though certainly not perfect, should 
somehow more realistically describe the evolutionary paths of such stars, since 
they~-- by definition~-- more successfully reproduce the internal structure of the 
Sun. 

To close, we note that \citet{vsp12} recently suggested that stellar metallicities 
derived in a way much as has just been described, based on helio- and asteroseismology, 
could be used to define a new, absolute stellar abundance scale. References 
to related prior work can be found in the reviews by 
\citet{mg09},
\citet{sb09,sb10},
and \citet{dg11,dg12}, among others.

\section{Mass Loss in Red Giants}\label{sec:m-dot}
Whether in their first or second ascent of the red giant branch, low-mass stars are expected
to lose large amounts of mass as they evolve. In the calculation of evolutionary tracks, 
it is still a common procedure to adopt the \citet{dr75a,dr75b} analytical expression to 
quantitatively evaluate the corresponding mass loss rates. However, many other 
formulations have been proposed in the more recent literature, and the differences 
in predicted quantities, such as the integrated RGB mass loss and its dependence on 
metallicity, can be quite substantial, with a strong impact on the predicted RGB progeny
\citep[e.g.,][and references therein]{mc09,jkea09}. It is thus very important to
establish proper recipes for the mass loss rates in low-mass red giants, guided by the 
latest available empirical data. 

In this sense, it is now well established that the \citeauthor{dr75a} 
expression does not properly describe 
the behavior of mass loss in red giants \citep[e.g.,][]{sc05,sc07,cs11}, and should thus 
not be used in state-of-the-art evolutionary calculations. 

Fortunately, several improved 
formulations have become available recently. In particular, \citet{sc05} proposed the 
following generalized formulation of the original \citeauthor{dr75a} expression
(their eq.~4, as modified by \cite{avalea12} for further 
flexibility in evolutionary computations): 

\begin{equation}
\dot{M} = 8 \times 10^{-14} \, \eta_{\rm SC} \,
   \frac{L_{\star} R_{\star}}{M_{\star}} \left(\frac{T_{\rm eff}}{4000~{\rm K}}\right)^{3.5}
   \left(1 + \frac{g_{\odot}}{4300 \, g_{\star}}\right)  \,\,\,\,\, (M_{\odot}\,{\rm yr}^{-1}), 
   \label{eq:sc05}
\end{equation}

\noindent where $L_{\star}$, $R_{\star}$, $M_{\star}$, are the stellar 
luminosity, radius, and mass, respectively, given in solar units; $T_{\rm eff}$ is 
the star's effective temperature, in K; $g_{\star}$ and $g_{\odot}$ are the stellar and
solar surface gravities, respectively; and $\eta_{\rm SC}$ is a dimensionless coefficient 
of order 1.  
In this way, as shown by \citet{sc05}, one is able 
to take into account both the mechanical energy flux that is added to the outer layers of 
the star and the dependence on chromospheric height, which are two key ingredients that 
were not satisfactorily included in the original \citeauthor{dr75a} formulation. 
The numerical factor $8 \times 10^{-14}$ was adjusted by 
\citet{sc05} to provide an optimum fit to HB stars in two GCs, 
namely M5 (NGC~5904) and NGC~5927. Then, assuming $\eta_{\rm SC} = 1$, 
\citet{sc07} have shown that, unlike with the \citeauthor{dr75a} formula, one is 
able to obtain good agreement, within the error bars, with the empirical data for well-studied 
red giants and supergiants. 

Very recently, a predictive theoretical model for the mass-loss rates of cool stars has 
been proposed by \citet{cs11} which seems to provide an even better fit to the empirical 
data \citep[see also][]{vaea10}. 
The model follows the flux of MHD turbulence from subsurface convection zones to its
eventual dissipation and escape through open magnetic flux tubes. The main drawback, as far
as applications to the calculation of stellar evolutionary tracks 
is concerned, is the fact that their formulation does not reduce to a 
simple analytical formula~-- no doubt the main reason for the continued popularity 
of the \citet{dr75a,dr75b} expression, its well-established deficiencies notwithstanding. 
Instead, information on each individual star's magnetic fields and activity is a priori 
required. 

Given the seemingly universal character of the quiet-Sun 
chromosphere among inactive stars, and thus the established concrete possibility of guiding 
mechanical energy toward the acceleration zone of cool stellar winds along flux tubes 
\citep{kpsea12}, further exploration of the \citet{cs11} formulation in actual 
evolutionary calculations would prove of great interest, and is thus strongly 
encouraged. In particular, magnetic field variations 
may play a relevant role in the context of the second-parameter problem of HB 
morphology \citep[see][for a review and references]{mc09a}, possibly being 
responsible, for instance, for the existence of star-to-star variations in total 
mass along the HBs of individual GCs \citep[e.g.,][]{vc08}, whose origin has long 
remained a mystery. 
In this sense, it would be especially interesting if the $\eta_{\rm SC}$
parameter in eq.~(\ref{eq:sc05}) could be calibrated in terms of the magnetic 
field-related quantities that appear in the \citeauthor{cs11} models.

\section{Multiple Populations in Globular Clusters}\label{sec:he}

It is now quite well established that GCs are not the simple stellar 
systems that they were once thought to be. In particular, evidence for the presence of 
multiple populations, interpreted as multiple star formation episodes with different 
degrees of associated chemical enrichment, has been obtained on the basis of both 
photometric and spectroscopic techniques \citep[see, e.g.,][for recent reviews and 
extensive references]{ar08,kb11,td11,vc11,rgea12}. 

One of the most exciting aspects brought about by these recent developments is the 
perceived possibility of finding a physically well-motivated 
solution to the long-standing second-parameter 
problem. Indeed, while it has long been suspected that the abundance anomalies that 
are observed in GC red giants could be associated with the morphology
of the HB \citep[e.g.,][]{jn81,cfp95}, it is only recently that large enough 
observational databases have been amassed that allow more reliable tests to be 
carried out. In this context, \citet{rgea10} argue that such 
abundance anomalies appear to be less important than only metallicity and age, as far 
as the second-parameter phenomenon of HB morphology goes. 

According to most authors \citep[e.g.,][]{dc08}, the main driver of blue HB 
morphologies, in the face of abundance variations in (mainly) O, Na, Mg, and Al, 
is the associated level of He enhancement that is brought about in the course of the 
operation of the Ne-Na and Mg-Al proton-capture cycles, which give rise, for instance, 
to the well-known O-Na anticorrelation. 
Uncertainties in the relevant nuclear reaction rates 
can be quite large, reaching orders of magnitude in some 
cases \citep[e.g.,][]{ciea10a,ciea10b,ciea11}. 
In fact, the precise level of 
He enhancement that may be associated with a given degree of O depletion/Na (and Al) 
enhancement is {\em not} known a priori. 
As an example, in Figure~\ref{fig:gc-chem} ({\em left panel}) 
the predicted correlation between [Na/Fe] and [O/Fe] \citep[adapted from][]{amea09} is 
shown for two well-known GCs, namely M13 (NGC~6205) and NGC~2808, both 
of which harbor large numbers of extreme HB stars. The {\em right panel}
in the same figure shows the associated trend between the helium abundance $Y$ and 
[O/Fe]. It is especially noteworthy that extremely low levels 
of [O/Fe] may be reached, with only a minor level of associated He enrichment. 
In addition, closely the same [O/Fe] and [Na/Fe] ratios may imply widely different 
He enrichment levels. Last but not least, 
note also that some of the variation in the observed abundance ratios 
may be evolutionary in nature (i.e., reflecting mixing
processes operating in the interiors of the red giants), as recently demonstrated by 
\cite{jp12}. It is important to keep these points 
in mind, before arriving at conclusions 
regarding the level of He enrichment that may be present   
in a given population, based solely on measurements of light-element abundance ratios.

\begin{figure}[t]
  \centerline{
  \includegraphics*[width=4.95in]{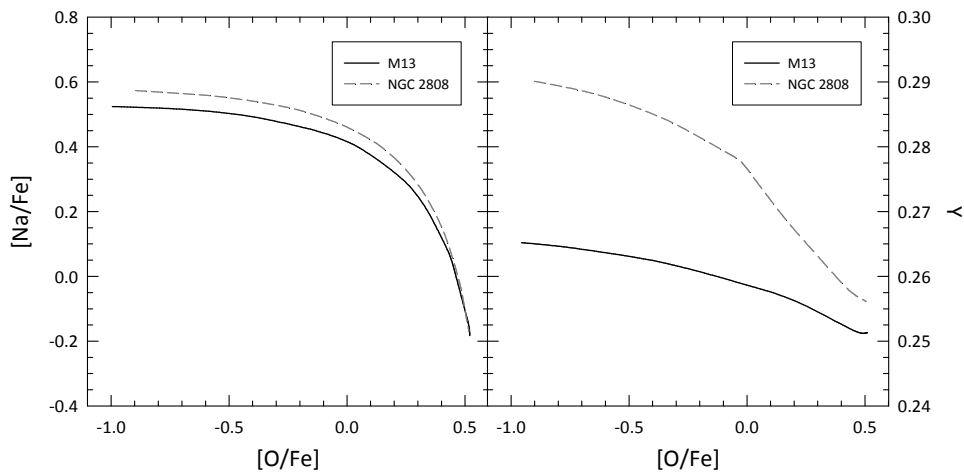}}
  \caption{Chemical evolution models for two GCs, namely M13 (NGC~6205, {\em solid
    lines}) and NGC~2808 ({\em dashed lines}). {\em Left:} the [Na/Fe] vs. [O/Fe] anticorrelation. 
	{\rm Right:} predicted correlation between helium abundance $Y$ and [O/Fe]. Note the different
	$Y$ levels that are predicted for the same [O/Fe]. Adapted from \citep{amea09}. 
   }
      \label{fig:gc-chem}
\end{figure}

As a case in point, consider the GCs M3 (NGC~5272) and M4 
(NGC~6121). As shown by \citet{csea04}, M3 possesses a large fraction of ``super-oxygen-poor'' 
stars, by which we mean stars with ${\rm [O/Fe]} < 0$. Based on this datum alone, one might 
be tempted to suspect that M3 possesses a large number of He-enhanced stars. Yet, as shown 
by \citet{mcea09} on the basis of high-precision Str\"omgren photometry, the level of 
He enhancement for the vast majority of M3 stars 
seems to satisfy $\Delta Y < 0.01$. M4, on the other hand, 
according to the spectroscopic measurements by \citet{amea08,amea11}, seems to be basically 
{\em devoid} of super-O-poor stars~-- and yet, according to \citet{svea12}, 
blue HB stars in 
this cluster somehow present a fairly high level of He enhancement, of order 
$\Delta Y \approx 0.02-0.04$. It remains to be established why it is that M3, with 
its significant levels of oxygen depletion, was somehow unable to produce HB stars 
with comparable levels of He enhancement as claimed for M4, 
whose stars do not reach similarly low [O/Fe] ratios as seen in M3.

\section{Conclusions}\label{sec:conc}
Several issues remain open in our knowledge of the evolutionary properties of low-mass 
stars. In this paper, a few topics that have been the subject of considerable debate in 
the recent literature have been addressed, namely the solar abundance problem, the mass 
loss rates in red giants, and the level of He enhancement in the different populations 
that are present within individual GCs. While considerable progress in all these areas 
has been made recently, more work will clearly be required before we are in a position 
to fully predict the evolutionary history of a low-mass star.  

\begin{acknowledgement} I would like to thank E. Caffau, H.-G. Ludwig, 
A. V. Sweigart, A. A. R. Valcarce, and N. Viaux, for several useful discussions and information. 
This work is supported by 
the Chilean Ministry for the Economy, Development, and Tourism's Programa 
Iniciativa Cient\'{i}fica Milenio through grant P07-021-F, awarded to The 
Milky Way Millennium Nucleus; by Proyecto Fondecyt Regular \#1110326; 
by the BASAL Center for Astrophysics and Associated Technologies 
(PFB-06); and by Proyecto Anillo ACT-86. 
\end{acknowledgement}

\end{document}